\documentclass[11pt, a4paper]{article}
\pdfoutput=1
\usepackage{jcappub}

\usepackage{booktabs}
\usepackage{multirow}
\usepackage{amssymb}
\usepackage[export]{adjustbox}
\usepackage{blindtext}
\usepackage{amsmath}
\usepackage{booktabs}
\usepackage{aas_macros}
\usepackage{soul}
\usepackage[normalem]{ulem}

\def\lsim{\mathrel{\raise.3ex\hbox{$<$\kern-.75em\lower1ex\hbox{$\sim$}}}}
\def\gsim{\mathrel{\raise.3ex\hbox{$>$\kern-.75em\lower1ex\hbox{$\sim$}}}}

\usepackage{appendix}
\usepackage{array}
\newcolumntype{x}[1]{>{\centering\arraybackslash}p{#1}}
\usepackage{bm}
\usepackage{color}
\usepackage{graphicx}
\usepackage{cancel}
\usepackage{amsfonts}
\usepackage{amssymb}
\usepackage{amsmath}
\usepackage{calc}
\usepackage{dcolumn}
\usepackage{mathrsfs}
\usepackage{mathtools}

\newcommand{\beq}{\begin{equation}}
\newcommand{\eeq}{\end{equation}}

\definecolor{rossoCP3}{cmyk}{0,.88,.77,.40}
\definecolor{verdeCP3}{rgb}{0.09765625, 0.57421875, 0.1015625}
\definecolor{bluCP3}{rgb}{0, 0.23, 0.67}

\ifx\pdfoutput\undefined
\usepackage[dvips,bookmarks]{hyperref}    
\else
\usepackage{hyperref}    
\fi
\hypersetup{colorlinks, bookmarksopen, bookmarksnumbered,
citecolor=verdeCP3, linkcolor=bluCP3, pdfstartview=FitH, urlcolor=rossoCP3}

\begin{document}

\vskip 0.2in

\title{Effect of axion-like particles on the spectrum of the extragalactic gamma-ray background}

\author{Yun-Feng Liang$^{a,}$\footnote{Corresponding author.},}
\emailAdd{liang-yf@foxmail.com}
\author{Xing-Fu Zhang$^a$,}
\emailAdd{zhangxf@st.gxu.edu.cn}
\author{Ji-Gui Cheng$^a$,}
\emailAdd{cheng-jg@qq.com}
\author{Hou-Dun Zeng$^b$,}
\emailAdd{zhd@pmo.ac.cn}
\author{Yi-Zhong Fan$^b$,}
\emailAdd{yzfan@pmo.ac.cn}
\author{En-Wei Liang$^{a,}$\footnote{Corresponding author.}}
\emailAdd{lew@gxu.edu.cn}

\affiliation[a]{Laboratory for Relativistic Astrophysics, Department of Physics, Guangxi University, Nanning 530004, China}
\affiliation[b]{Key Laboratory of Dark Matter and Space Astronomy, Purple Mountain Observatory, Chinese Academy of Sciences, Nanjing 210008, China}

\date{\today}

\abstract{
Axion-like particles (ALPs) provide a feasible explanation for the observed lower TeV opacity of the Universe. If the anomaly TeV transparency is caused by ALPs, then the fluxes of distant extragalactic sources will be enhanced at photon energies beyond TeV, resulting in an enhancement of the observed extragalactic gamma-ray background (EGB) spectrum. In this work, we have investigated the ALP modulation on the EGB spectrum at TeV energies. Our results show that in the most optimistic case, the existence of ALPs can cause the EGB spectrum to greatly deviate from the prediction of a pure extragalactic-background-light (EBL) absorption scenario. The deviation occurs at approximately $\gtrsim$1 TeV, and the current EGB measurements by Fermi-LAT cannot identify such an effect. We also find that most of the sensitive ALP parameters have been ruled out by existing constraints, leaving only a small region of unrestricted parameters that can be probed using the EGB effect investigated in this work. Observations from forthcoming very-high-energy instruments like LHAASO and CTA may be beneficial for the study of this effect.
}

\maketitle

\section{Introduction} \label{sec:intro}
Axion-like particles (ALPs) are hypothetical particles predicted by several extensions of the Standard Model \cite{rv1,rv2}. They can also be candidates for dark matter (DM) in the universe. One property predicted by the ALP model is that ALPs can interact with photons in an external magnetic field. The interaction is described by the Langrangian, $\mathcal{L}=g_{a \gamma} \vec{E} \cdot \vec{B} a$, where $g_{a\gamma}$ is the ALP-photon coupling, $a$ is the ALP field and $B$ is the external magnetic field.
The resulting astrophysical effects due to the ALP-photon interaction have been widely studied in recent years. One effect is that the conversion between ALPs and photons offers a possible explanation for the low observed opacity of the Universe for TeV photons \cite{mirizzi7,angelis07,MirizziA2009,Sanchez-Conde09,BelikovAV2011,HornsD2012a,MeyerM2014b,ReesmanR2014,Kohri17,LongGB2020}.

Very-high-energy (VHE, $>100\,{\rm GeV}$) gamma rays from distant sources interact with environmental photons during their propagation, and convert into $e^+e^-$ pairs, which prevent them from propagating a long distance and arriving at the Earth.
The determination of the TeV optical depth ($\tau_{\rm TeV}$) relies on the observation and modeling of the energy density of the extragalactic background light (EBL). Though the EBL is not exactly known, the minimum EBL model can be derived from galaxy number counts \cite{ebl_Kneiske}.
In the literature, we have evidence that $\tau_{\rm TeV}$ is lower than (or at least very close to) the minimum model prediction \cite{Protheroe2000,magic08,3c66a_magic,3c66a_veritas,Angelis09_tev,dominguez2011alp,HornsD2012b,RubtsovGI2014,Kohri2017}, suggesting that possibly an additional effect makes the universe more transparent.

The anomaly TeV transparency can be explained reasonably in the ALP framework \cite{mirizzi7,angelis07,MirizziA2009,Sanchez-Conde09,BelikovAV2011,HornsD2012a,MeyerM2014b,ReesmanR2014,Kohri17,LongGB2020}. In the magnetic fields near the source, VHE photons may be converted into ALPs, which can travel unimpeded in space, avoiding any interaction with background photons.
The reconversion of ALPs back into VHE photons in the Milky Way's magnetic field makes them detectable by us. 
It has been shown that for 1ES 0414+009 and Mkn 501, up to 10\% of the emitted flux can survive because of this effect \cite{HornsD2012b}.
\citet{Meyer2013} derived the corresponding parameter region that can account for the TeV transparency issue with the ALP effect.
Part of this region has been ruled out by the Fermi-LAT observation of NGC 1275 \cite{AjelloM2016,AdamanePallathadkaG2020,cjg2020} and the HESS observation of PKS 2155-304 \cite{Abramowski2013}. This region will be further probed by the future CTA telescope \cite{MeyerM2014a,LiangYF2019,cta_alp}. We would like to caution that the ALP effect isn't the only explanation for the lower opacity, and some other works suggest that the current VHE observations are consistent with EBL-only expectations \cite{Sanchez2013,Biteau2015,dominguez2015,Meyer16proc}. 
Ref. \cite{BelikovAV2011} used the observations of 1ES1101-232 and H2356-309 to test the ALP hypothesis and find that ALPs are not necessary to explain the observed degree of attenuation. 
Ref. \cite{Buehler2020} does not ﬁnd any hint of the existence of ALPs in their search for enhanced transparency using Fermi-LAT data and set limits on the ALP parameters.

Though the ALP effects in the VHE range have been widely studied (see also \cite{XiaZQ2019,BiXJ2020,lihaijun2020} for other works on ALPs at TeV energies for both galactic and extragalactic sources), 
all the above-mentioned works are based on resolved TeV sources beyond the detection threshold. In this paper, we show that the ALP-photon conversion can also significantly modify the spectrum of the diffuse extragalactic gamma-ray background (EGB) in the $>100\,{\rm GeV}$ energy range.
This effect has not been studied in the previous works yet. 
{As the VHE measurements of EGB are still absent, we mainly aim to assess the maximal effect ALPs could have on the EGB spectrum rather than give a constraint.}
We point out that such modulation on the EGB may be identified by space-based/ground gamma-ray telescopes if the magnetic fields near the sources are strong enough.

\section{Calculations}

\subsection{EGB spectrum} \label{sec:egrb}
The EGB represents all observed gamma-ray emissions from both resolved and unresolved sources outside the Milky Way. The EGB spectrum from 0.1 GeV to 800 GeV has been well measured by the Fermi-LAT \citep{fermi15iso}. This spectrum is best fitted with a power law with exponential cut-off; the best-fit photon index and cut-off energy are $\sim$2.3 and $\sim300$ GeV, respectively \citep{fermi15iso}. 
The cut-off at high energies is mainly caused by EBL absorption.
Previous analyses have shown that although star-forming galaxies and radio galaxies contribute a lot to the EGB at lower energies, at energies of $>50\,{\rm GeV}$ the gamma-ray blazars can account for almost the totality of the EGB \citep{ajello15egb,zeng_ebl}. 
As the blazar population can account for all the $>50$ GeV EGB, any effect (e.g. ALP) leading to further enhancement of the spectrum will not be favoured by the observation.

Based on the resolved blazars detected by Fermi-LAT and their redshift information, the luminosity function (LF) of the whole blazar population can be inferred. Thus, the contribution to the EGB by blazars (including unresolved blazars) can be calculated. The LF $\Phi\left(L_{\gamma}, z, \Gamma\right)$ is defined as the space number density of blazars as a function of rest-frame
0.1–100 GeV luminosity ($L_{\gamma}$), redshift ($z$) and photon index ($\Gamma$).
Here, we use the formulae and parameters in \cite{ajello15egb} to compute the EGB spectrum contributed by blazars,
\begin{eqnarray}
F_{\rm EGB}(E_{\gamma})&=&\int\limits_{\Gamma_{\min}=1.0}^{\Gamma_{\max}=3.5} d\Gamma \int\limits_{z_{\min}=10^{-3}}^{z_{\max}=6} dz \nonumber \\
&\times&\int\limits_{L_{\gamma}^{\min}=10^{43}}^{L_{\gamma}^{\max}=10^{52}} dL_{\gamma} \cdot \Phi\left(L_{\gamma}, z, \Gamma\right) \cdot \frac{d N_{\gamma}}{d E} \cdot \frac{d V}{d zd\Omega} \nonumber \\
&\times&\left({\rm ph\,cm^{-2}s^{-1}sr^{-1}GeV^{-1}}\right)
\label{eq:egbsp}
\end{eqnarray}
where ${dV}/{dz/d\Omega}$ is the comoving volume element per unit redshift per unit of the solid angle. We use the cosmological parameters of \cite{planck15}. The LF is also rescaled with the updated cosmological parameters following \cite{zeng_ebl}. For each blazar, its spectrum is assumed to be 
\begin{eqnarray}
\frac{d N_{\gamma}}{d E}=K\left[\left(\frac{E}{E_{b}}\right)^{\gamma_{a}}+\left(\frac{E}{E_{b}}\right)^{\gamma_{b}}\right]^{-1} \cdot P_{\gamma\gamma}(E,z) \nonumber \\
\quad\times\left({\rm ph\,cm^{-2}{s}^{-1}GeV^{-1}}\right)
\label{eq:dnde}
\end{eqnarray}
where $P_{\gamma\gamma}$ is the modulation factor due to the EBL and ALP effects. For EBL absorption, it is $P_{\gamma\gamma}=e^{-\tau(E, z)}$ (see Section \ref{sec:od}) and for ALP it is the photon survival probability $P_{\gamma\gamma,\rm ALP}$ in Section \ref{sec:alp}. By modifying the traditional $e^{-\tau}$ factor by $P_{\gamma\gamma,\rm ALP}$, we obtain the expected EGB spectrum in the ALP scenario.

\subsection{Optical depth for TeV photons}
\label{sec:od}

TeV photons undergo absorption due to the interaction with low energy EBL photons. The absorption rate as a function of observed energy $E$ and source redshift $z$ is 
\begin{equation}
\Gamma(E,z)=\int_{0}^{2}{\rm d}\mu\frac{\mu}{2}\int_{\epsilon_{\rm th}}^\infty{\rm d}\epsilon' \sigma_{\gamma\gamma}(\beta')n(\epsilon',z)
\end{equation}
where $\epsilon$ and $n(\epsilon,z)$ are the EBL photon energy and number density, $\sigma_{\gamma\gamma}$ is the photon-photon pair production cross section. The energy threshold of the pair production interaction is
$\epsilon_{th}={2m_{\rm e}^2c^4}/(E'\mu)$, where $\mu=(1-\cos \theta)$ with $\theta$ the the angle of the interaction. The $\beta'$ is
\begin{equation}
\beta'=\frac{\epsilon_{th}}{\epsilon'(1+z)^2}.
\end{equation}
A prime means the quantity is in the rest-frame at the redshift of the source.

The optical depth of the gamma-ray photons emitted by extragalactic objects is thus
\begin{eqnarray}
\tau_{\gamma\gamma}(E,z)&=&\int_{0}^{z}\Gamma(E,z')\left(\frac{{\rm d}l'}{\mathrm{d} z'}\right){\rm d}z'\\
&=& \frac{c}{H_0}\int_{0}^{z}\frac{dz'\Gamma(E, z')}{(1+z') \sqrt{\Omega_{\Lambda}+\Omega_{\mathrm{m}}(1+z')^{3}}}.
\end{eqnarray}
The EBL in the optical to infrared range is primarily contributed by the light of galaxies through the evolutionary history of the Universe. Several EBL models are available in the literature (e.g. \cite{Franceschini08,ebl_Kneiske,finke2010,DominguezA2011,Inoue13_ebl}). In this work, we use the model from \cite{DominguezA2011} as a benchmark EBL model\footnote{The EBL and optical depth data are publicly available at \url{http://side.iaa.es/EBL/}.}.

\subsection{ALP effect}
\label{sec:alp}
In the ALP scenario, photons can be converted into ALPs in the magnetic field near the source and the latter will travel through extragalactic space unimpeded. 
The photon survival probability after propagation can be computed by solving the propagation equation of the ALP-photon beam. A number of previous articles have elaborated the calculation procedure, and we mainly refer to \cite{MeyerM2014a} and the references therein. 
The code we used to calculate the ALP-photon conversion is available online\footnote{\url{https://github.com/lyf222/alpconv}}. 
We have double-checked the major calculation steps with the publicly available code {\tt gammaALPs}\footnote{\url{https://gammaalps.readthedocs.io/en/latest/index.html}} \cite{2021arXiv210802061M} and find that the two codes give consistent results.

The propagation equation for a photon-ALP beam is (propagating along $x_3$ direction)
\begin{equation}
\label{eq:propa}
    \left(i\frac{\mathrm{d}}{{\mathrm{d}x}_{3}}+E+\mathcal{M}_0\right)\Psi(x_{3})=0
\end{equation}
where $\Psi(x_{3})$ is the state function of the photon-ALP beam and $\mathcal{M}_0$ is the photon-ALP mixing matrix
\begin{equation}
    \mathcal{M}_0 = \left(
        \begin{array}{ccc}
        \Delta_{\perp}-\frac{i}{2}\Gamma & 0 & 0 \\
        0 & \Delta_{\parallel}-\frac{i}{2}\Gamma & \Delta_{a\gamma} \\
        0 & \Delta_{a\gamma} & \Delta_{a}
        \end{array}
    \right)
\label{eq:matrix}
\end{equation}
The elements $\Delta_{\perp}$, $\Delta_{\parallel}$, $\Delta_{a}$, $\Delta_{a\gamma}$ depend on ALP mass $m_a$, coupling $g_{a\gamma}$, photon energy $E$, strength of the transverse magnetic field $B_{\rm T}$ and electron density $n_{\rm e}$, which reads \citep{MirizziA2009}
\begin{eqnarray}
\Delta_{a \gamma} & \simeq& 1.52 \times 10^{-2}\left(\frac{g_{a \gamma}}{10^{-11} \mathrm{GeV}^{-1}}\right)\left(\frac{B_{T}}{\mu \mathrm{G}}\right) \mathrm{kpc}^{-1} \nonumber\\
\Delta_{a} & \simeq&-7.8 \times 10^{-3}\left(\frac{m_{a}}{10^{-8} \mathrm{eV}}\right)^{2}\left(\frac{E}{\mathrm{TeV}}\right)^{-1} \mathrm{kpc}^{-1} \nonumber\\
\Delta_{\mathrm{pl}} & \simeq&-1.1 \times 10^{-10}\left(\frac{E}{\mathrm{TeV}}\right)^{-1}\left(\frac{n_{\rm e}}{10^{-3} \mathrm{~cm}^{-3}}\right) \mathrm{kpc}^{-1} \nonumber\\
\Delta_{\mathrm{QED}} & \simeq& 4.1 \times 10^{-6}\left(\frac{E}{\mathrm{TeV}}\right)\left(\frac{B_{T}}{\mu \mathrm{G}}\right)^{2} \mathrm{kpc}^{-1},
\end{eqnarray}
\begin{eqnarray}
\Delta_{\perp}&=\Delta_{\mathrm{pl}}+2 \Delta_{\mathrm{QED}}+\Delta_{\gamma \gamma} \nonumber\\
\Delta_{\|}&=\Delta_{\mathrm{pl}}+\frac{7}{2} \Delta_{\mathrm{QED}}+\Delta_{\gamma \gamma}.
\end{eqnarray}
The photon-photon dispersion term $\Delta_{\gamma\gamma}$ is introduced to consider forward scattering on real photons below the pair-production threshold  (dominated by the CMB), which is especially relevant for the VHE emission from extragalactic sources \cite{Dobrynina2015,Kartavtsev2017}.
The EBL absorption is taken into account by including an additional term $-{i\Gamma}/{2}$ in the mixing matrix (Eq. (\ref{eq:matrix})), where $\Gamma$ is the EBL absorption rate described in Section \ref{sec:od}. For energies beyond a few tens of TeV, the absorption by the interstellar radiation ﬁeld (ISRF) of the Milky Way becomes important \cite{VogelH2017}. We also include the ISRF absorption in the calculation \cite{Moskalenko2006,Popescu2017}. It is, however, found to be neglectable since it reduces the flux only at the $\sim10\%$ level at the energy of 100 TeV for the emission off the Galactic plane.

By solving Eq.(\ref{eq:propa}) we can derive the density matrix 
\begin{equation}
    \rho(x_{3})=\Psi(x_{3})\Psi(x_{3})^{\dagger}=\mathcal{T}(E,x_3)\rho(0)\mathcal{T}^{\dagger}(E,x_3)
\end{equation}
with $\mathcal{T}$ the full transfer matrix (the explicit form can be found in \cite{Sanchez-Conde09,BassanN2010}). 

The final photon survival probability is given by
\begin{equation}
\label{eq:survial_prob}
    P_{\gamma\gamma}(E,x_3)=\mathrm{Tr}((\rho_{11}+\rho_{22})\mathcal{T}(E,x_3)\rho(0)\mathcal{T}^{\dagger}(E,x_3))
\end{equation}
where $\rho_{11}={\rm diag}(1,0,0)$, $\rho_{22}={\rm diag}(0,1,0)$.

\subsection{Magnetic field environments}
\label{sec:mf}
\begin{figure*}[t]
	\centering
	\includegraphics[width=0.49\textwidth]{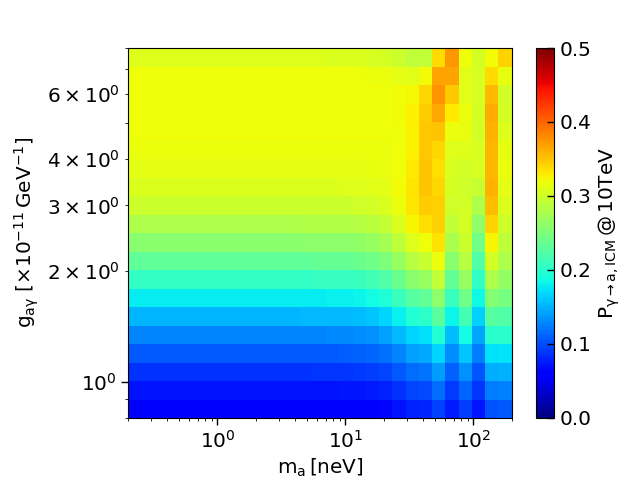}
	\includegraphics[width=0.49\textwidth]{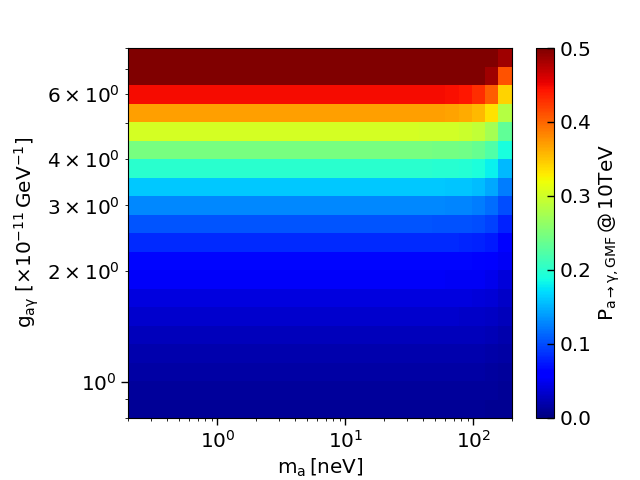}
	\caption{The probability of photons converting into ALPs in ICM ($P_{\gamma\rightarrow a}$, left panel) and ALPs converting back into photons in GMF ($P_{a\rightarrow\gamma}$, right panel) as a function of ALP mass $m_a$ and ALP-photon coupling $g_{a\gamma}$. We show the conversion probability at the photon/ALP energy of 10 TeV. A higher $P_{\gamma\rightarrow a,{\rm ICM}}$ or $P_{a\rightarrow\gamma,\rm GMF}$ will lead to a greater increase of the TeV transparency. These two plots show that the $\gamma\rightarrow a$ conversion is efficient even for $g_{a\gamma}<2\times10^{-11}\,{\rm GeV^{-1}}$, while for ${a\rightarrow\gamma}$ the conversion is only efficient when $g_{a\gamma}>4\times10^{-11}\,{\rm GeV^{-1}}$, indicating for the given ICM configuration the ALP effect on EGB spectrum is mainly limited by the GMF.}
	\label{fig:pmap}
\end{figure*}

To avoid the EBL absorption and have a stronger ALP effect, it is required more photons are converted into ALPs in the magnetic field close to the source. Simultaneously, these ALPs must be converted back to photons in the magnetic field near the observer, so that they can be detected. In this work, we consider the combination of the intracluster magnetic field (ICM) and the magnetic field of the Milky Way (GMF). It has been shown that this combination can effectively reduce the TeV opacity and enhance the VHE flux (e.g. \cite{Sanchez-Conde09,HornsD2012b,Meyer2013,MeyerM2014b,WoutersD2014,GuoJG2020}). For ICM, the magnetic field environments in the galaxy clusters around different blazars are not the same. Faraday rotation measurements show that the strengths are between 1 and 10 ${\rm \mu G}$. For a demonstration purpose, we use the same ICM environment for all blazars  as that of NGC 1275 adopted in \cite{AjelloM2016} (but the $B_0$ is set to a more moderate value of $3\mu G$). 
We will see below that the ALP effect on EGB spectrum is mainly limited by the GMF (see Figure \ref{fig:pmap}).

It is pointed out that the ICM adopted in \cite{AjelloM2016} can not reproduce the observed Faraday rotation measure of NGC 1275, and a regular field component should be taken into account, which will significantly reduce the irregularity caused by ALP-photon conversion \citep{Libanov2020}. However, for a magnetic-field region of $\sim90\,{\rm kpc}$ with a mean strength of $\sim3\,\mu{\rm G}$ \cite{Libanov2020}, we still have $15\times({g_{a\gamma}}/{10^{-11}{\rm GeV}^{-1}}) (B/{\rm G})(L/{\rm pc}) \gtrsim 1$ \citep{WoutersD2014} below the CAST limit, indicating that the conversion in the strong mixing regime remains efficient (also see \cite{cjg2020}). Furthermore, the photons can be converted into ALPs in the magnetic field of a blazar jet as well \citep{alpagnjet}, providing another environment for ALP-photon conversion near the source. 

{We'd like to emphasize that, the above ICM model assumes that all blazars are located within galaxy clusters, which is an overly optimistic assumption. The realistic environment surrounding AGNs would be complex and very uncertain. Along with the fact that the EGB observations at relevant energies are still absent yet. In this work, our primary goal is not to provide robust constraints on ALP parameters, but rather to demonstrate how much the EGB spectrum will be maximally modulated by the ALP effect.}

For GMF, we consider the model of \cite{JanssonR2012} that can best fit the observations of Faraday rotation measures and polarized synchrotron radiation. This model incorporates an additional out-of-plane component, which is supported by the observations of external edge-on galaxies and guarantees a relatively high photon-ALP conversion at high latitudes. The turbulent component of GMF is ignored since its coherence length is much smaller than the photon-ALP oscillation length. 
It's worth noting that the GMF is anisotropic in the sky, hence the $P_{\gamma\gamma}$ is direction-dependent. Precise prediction necessitates computing the $P_{\gamma\gamma}$ (as a function of energy $E$ and redshift $z$) for every direction of the entire sky, which is time-consuming. For a demonstration purpose, here we use the $P_{\gamma\gamma}$ calculated with the coordinates of NGC1275 as a proxy of the allsky-averaged value.
In light of its coordinates of $l=150.6^\circ,b=-13.3^\circ$, it is a somewhat conservative choice \cite{WoutersD2014}. We put the source at different redshifts to compute the photon survival probability as a function of energy $E$ and redshift $z$, $P_{\gamma\gamma,\rm ALP}(E,z)$.

{For the intergalactic magnetic field (IGMF), its strength remains less constrained. A generally accepted upper limit on the Mpc scale is below $<10^{-9}\,{\rm G}$ \cite{Blasi99igmf,Pshirkov16igmf,Bray18uhecr} and the galaxy cluster simulation gives $B_{\rm IGMF}\sim10^{-12}\,{\rm G}$ \cite{Neronov10sci}.
Very small IGMF values have also been advocated in the literature (e.g. $B_{\rm IGMF}\sim0$) \cite{Arlen14_0igmf,chen15agnhalo}, though they are not supported by the lower limits derived from the non-detection of electromagnetic halos in the Fermi-LAT observation of TeV blazars \cite{fermi18ext}. 
A recent analysis of the cosmic-microwave background constrains the magnetic field to be $<0.047\,{\rm nG}$ \cite{Jedamzik19igmfPRL}.
Even for the largest redshift of $z=6$ in the integration of Eq. (\ref{eq:egbsp}), we find that the photon-ALP conversion is only meaningful with an IGMF higher than $\sim10^{-10}\,{\rm nG}$. 
Due to the substantial uncertainty of the IGMF and the possible very low values, we do not consider the IGMF component in our analysis.}

\section{Results and discussions}

\begin{figure}[th]
	\centering
	\includegraphics[width=0.6\textwidth]{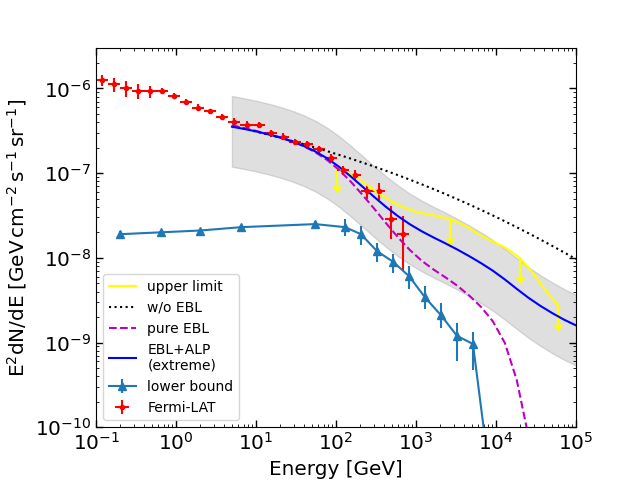}
	\caption{The EGB spectra observed by Fermi-LAT (red points) and calculated with the blazar luminosity function (dotted, solid and dashed lines). The dotted, dashed and solid lines represent the EGB spectra of no EBL absorption (w/oEBL), with EBL considered but no ALP effect (EBL) and including an ALP effect (ALP), respectively. We have assume an optimistic ICM for the ALP model. The gray band reflects the scatter caused by the uncertainty of the LF parameters. The triangles are the lower limits of the EGB spectrum placed by cumulating detected individual extragalactic TeV sources \cite{Inoue2016}. The yellow line is the upper limit on EGB by requiring the cascade emission not to exceed the EGB data below 100 GeV \cite{Inoue2012}.}
	\label{fig:egb}
\end{figure}

\begin{figure}[th]
	\centering
	\includegraphics[width=0.49\textwidth]{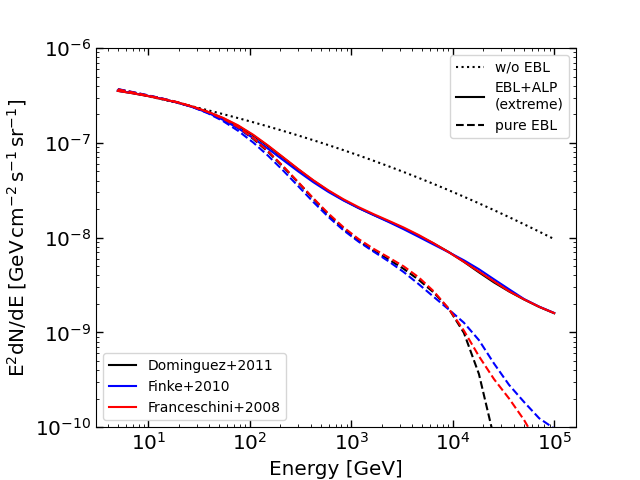}
	\includegraphics[width=0.49\textwidth]{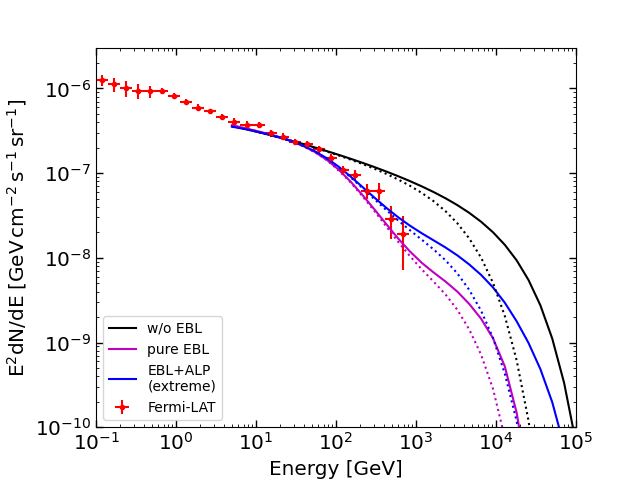}
	\caption{({\it left panel}) The model expected EGB spectra based on different EBL models. ({\it right panel}) The EGB spectra of considering a high energy cutoff $E_{\rm c}=20\,{\rm TeV}$ (solid lines) or $E_{\rm c}=5\,{\rm TeV}$ (dotted lines) in the intrinsic blazar spectrum. }
	\label{fig:ebl}
\end{figure}

We use the equations mentioned in the previous section to calculate the model-expected EGB spectrum, with the goal of demonstrating how the EGB spectrum will vary in the ALP scenario.
We compute the contribution to the EGB from blazars using the pure density evolution (PDE) LF in \cite{ajello15egb}. The parameters in Table 1 of \cite{ajello15egb} are used. We select the ALP parameters of $m_a=1\,{\rm neV},\ g_{a\gamma}=5\times10^{-11}\,{\rm GeV^{-1}}$ to demonstrate the effect. Figure \ref{fig:egb} depicts the results. The dotted, dashed and solid lines represent the EGB spectra of no EBL absorption (w/oEBL), with EBL considered but no ALP effect (EBL) and including an ALP effect (ALP), respectively. As shown, when the ALP effect is taken into account, the model-expected EGB spectrum begins to deviate from the other two at energies greater than $1$ TeV.
The gray band in Figure \ref{fig:egb} represents the scatter due to the uncertainty of the LF parameters.
In addition, the choice of EBL models also contributes to the uncertainty in EGB calculation.
In Figure \ref{fig:ebl}, we demonstrate the expected spectra for various EBL models. We take a look at three EBL models, i.e. \citet{Franceschini08,finke2010,DominguezA2011}. The EBL model is found to have only a minor effect on the results.

\begin{figure}[t]
	\centering
	\includegraphics[width=0.6\textwidth]{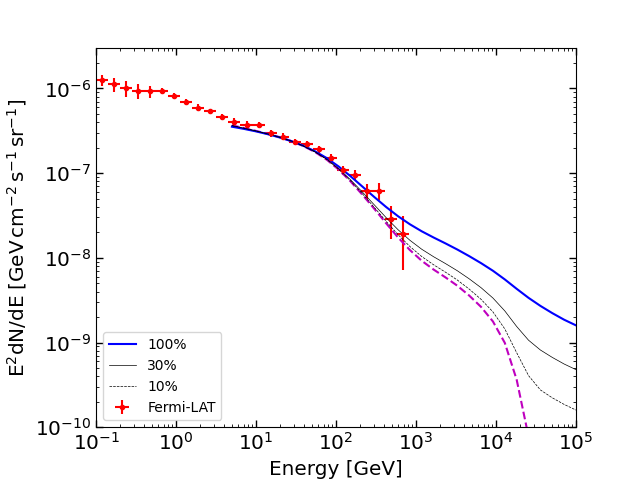}
	\caption{The modulated EGB spectra assuming only 10\% (grey solid line) or 30\% (grey dotted line) of blazars are located in a dense intracluster environment.}
	\label{fig:10p}
\end{figure}

For the ICM, the above results assume that all blazars contributing to the EGB are located in the same environment, i.e., the same magnetic ﬁeld as in the Perseus galaxy cluster but with a lower central value of $3\,{\rm \mu G}$ ({extreme ICM}). The results shown in Figure \ref{fig:egb} therefore represent the most optimistic case. The realistic magnetic field environments around AGNs would be complex and very uncertain, e.g., some blazars are in galaxy clusters and some others isolated. Taking this into account, we test how the results will change if not all blazars are located in such a dense environment. We use a simple way to demonstrate the influence. We simply suppose that only 10\% (or 30\%) of AGNs are located in clusters with a magnetic field environment similar to {the extreme ICM}, and the rest are isolated AGNs of which the emission does not pass through any ICMs (or the ICMs are weak and ignorable). The results are shown in Figure \ref{fig:10p}. We find that if the majority of the AGNs are not located in an `{extreme ICM}' environment, the ALP effect discussed in this work can barely affect the EGB spectrum. 
The deviation between the ALP and pure EBL models is only significant up to $\sim100\,{\rm TeV}$. Together with the possible cutoff of the spectrum at high energies (see below), the effect is therefore difficult to detect in a more realistic scenario. However, as previously stated, the emission from a blazar will always pass through a jet magnetic field, which provides another site for photon-ALP conversion near the source \citep{alpagnjet}. We leave related studies considering the jet magnetic field in future works.

In the calculation, we assume a power-law extrapolation of the blazar's intrinsic spectrum from tens of GeV to 100 TeV  (Eq. (\ref{eq:dnde})). {The real spectrum is likely softer at the high energy end (outside of the Fermi-LAT energy range). Due to the maximum energy of the accelerated particles or pair interaction in the source region, there may be a high energy cutoff in the spectrum. Here we test how will the results change if we impose lower cut-off energies on the spectrum. This is implemented by multiplying an extra $\exp({-E/E_c})$ term in Eq. (\ref{eq:dnde}). We adopt two $E_{\rm c}$ in the calculation, $E_{\rm c}=5\,{\rm TeV}$ and $20\,{\rm TeV}$. The results are plotted in the right panel of Fig. \ref{fig:ebl}. 
As shown, if the spectrum is cut off at lower energies, the deviation between ALP and pure EBL models becomes less substantial (especially for the case of $E_{\rm c}=5\,{\rm TeV}$), making it difficult to identify the ALP effect even when measuring the EGB at 10 TeV.} Nevertheless, blazars may have extra VHE components (from $p\gamma$ or $pp$ interaction) at high energies.
Regardless of coming from the $p\gamma$ or $pp$ mechanism, VHE photons will be produced alongside the production of neutrinos. 
The observed high diffuse neutrino flux \citep{icecube2020,Murase15neutrino} indicates a high intrinsic photon flux. Therefore, the power-law extrapolation to 100 TeV is optimistic but still reasonable.

Another issue is that, even in the optimistic scenario, measuring the EGB at $\gtrsim1$ TeV energies is difficult. In Figure \ref{fig:egb}, also plotted are the Fermi-LAT measurements of the EGB \citep{ajello15egb}. We can see that the Fermi-LAT observation is insufficient to distinguish between models (wALP and w/oALP), since the deviation occurs at energies of greater than a few TeV (outside the sensitive range of Fermi-LAT). 
A lower bound on the EGB at TeV energies can be derived from the cumulative low-state flux of known extragalactic TeV sources \cite{Inoue2016} and an upper limit can be obtained by requiring the cascade emission not to exceed the EGB data below 100 GeV \cite{Inoue2012}. However, these results are not able to provide any hints since they are below/above both the EBL and APL models. The DAMPE satellite has a larger BGO calorimeter and is capable of detecting gamma rays up to 10 TeV \citep{dampe}. 
It also has much better cosmic-ray background rejection for photon data, which is critical for extracting the EGB spectrum. On the other hand, the acceptance for DAMPE is smaller (the effective area is $\sim10\%$ and field of view is $\sim1/2.4$ of Fermi-LAT). According to a simple estimation, 10 years of DAMPE observations detect $<1$ photons around 10 TeV even for the ALP model.

\begin{figure}[t]
	\centering
	\includegraphics[width=0.6\textwidth]{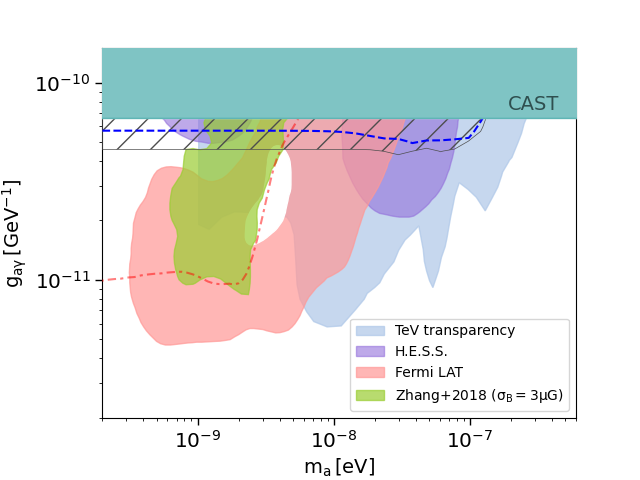}
	\caption{The ALP parameters sensitive to EGB observations (hatched region for $B_0=3\,{\rm \mu G}$ and blue dashed line for $B_0=1\,{\rm \mu G}$) comparing to other ALP constraints in the literature. The purple, red and green regions are the constraints by \citet{Abramowski2013}, \citet{AjelloM2016} and \citet{ZhangC2018}, respectively. {The red dash-dot line encompasses the exclusion region by \citet{Buehler2020}.} The light blue region is the parameter space where the low gamma-ray opacity of the universe can be explained by the ALPs \citep{Meyer2013}.}
	\label{fig:fig5}
\end{figure}

Ground-based Cherenkov telescopes 
like the Large High Altitude Air Shower Observatory (LHAASO) \citep{lhaaso} and the Cherenkov Telescope Array (CTA) \citep{cta} can observe photons up to $>100\,{\rm TeV}$ with very high statistics (i.e. very small statistical uncertainty), thus may be able to distinguish the models.
The problem for Cherenkov telescopes is that they do not have the ability to separate cosmic-ray electrons from gamma rays. 
The shower morphology only allows the instrument to distinguish between leptonic and hadronic cosmic rays.
For an isotropic signal, the on-/off-source analysis routinely employed by Cherenkov telescopes to identify sources cannot be used, either.
However, if the ALP-caused diffuse signal is not isotropic due to the anisotropic magnetic structure of MW, as suggested in \citet{VogelH2017}, morphology information then can be used to identify the ALP signal.

A larger future space-based telescope may be helpful for the study of the effect discussed in this work. Many gamma-ray satellites have been proposed in recent years, e.g. GAMMA-400, HERD and VLAST. 
We estimate that a 10-year observation of a next-generation telescope with an effective area 5 times greater than Fermi-LAT can detect $>$40 EGB photons around 10 TeV for the ALP model, enough to determine the EGB models at a $>5\sigma$ confidence level.
It should be noted that the estimation here only takes into account the statistical uncertainty due to Poisson fluctuation, which does not capture all of the expected scatter of spectral measurements. The uncertainty also comes from the procedure of disentangling the EGB from other emission components.

If the observed EGB spectrum does not exhibit any sign of a low optical depth (i.e. the observed EGB is consistent with the pure EBL model), the ALP parameter space will be constrained {(under the optimistic scenario)}. 
According to the photon survival probability we estimate the parameters that may cause the effect discussed in this work. 
We show the results in Figure \ref{fig:fig5}.
{Also shown are the constraints from \citet{Abramowski2013}, \citet{AjelloM2016} and \citet{ZhangC2018}. The red dash-dot line encompasses the exclusion region by \citet{Buehler2020}, which also concerned on the spectral characteristics at high energies for the blazar population. 
For ALP masses $m_a\lesssim3\,{\rm nG}$, they obtain stronger limits assuming an IGMF strength of 1 nG. Below $m_a\sim 1\,{\rm neV}$, the parameters of $g_{a\gamma}\gtrsim6\times10^{-12}\,{\rm GeV}$ are also constrained by the non-observation of accompanied gamma-ray signals from nearby supernova (SN) explosions due to the ALPs generated in the SN core \cite{Payez15sn1987,Meyer20sne}.}
As demonstrated, existing observations have ruled out most of the sensitive parameters (hatched region), leaving only a very few unconstrained parameters that can be probed based on the EGB observation.

Varying the $B_0$ of the ICM, which is the parameter that has the biggest influence on the results \cite{ZhangC2018},  to $1\,{\rm \mu G}$ yields the blue dashed line result.
In the above, we use the GMF in the direction of NGC 1275 to carry out the computations. As mentioned in Section \ref{sec:alp}, this is a somewhat conservative choice. 
If we consider EGB observations in other directions (e.g. a region around $l=0^\circ,b=30^\circ$) where GMF is stronger, the sensitive ALP parameters will be further improved to $\sim3\times10^{-11}\,{\rm GeV^{-1}}$.
It should be noted that these results rely on EGB observations at 10 TeV energy, which will only be achieved by future gamma-ray telescopes.

\begin{acknowledgments}
We thank the anonymous referee for the evaluation of our paper. 
We thank Jin Zhang and Nenghui Liao for help discussions.
This work is supported by the National Natural Science Foundation of China (Nos. 11851304, U1738136, 11533003, U1938106, 11703094) and the Guangxi Science Foundation (2017AD22006,2019AC20334,
    2018GXNSFDA281033) and special funding for Guangxi distinguished professors.
\end{acknowledgments}

\bibliographystyle{apsrev4-1-lyf}
\bibliography{references}

\end{document}